\documentclass{cs19proc}

\usepackage{kantlipsum}
\usepackage[breaklinks=true]{hyperref}
\hypersetup{
    bookmarksopen=false,
    bookmarksnumbered=false,
    unicode=true,           
    pdftoolbar=true,        
    pdfmenubar=true,        
    pdffitwindow=false,     
    pdfstartview={FitH},    
    pdfnewwindow=true,      
    colorlinks=true,       
    linkcolor=red,          
    citecolor=magenta,        
    filecolor=green,      
    urlcolor=blue           
}

\editors{G.~A. Feiden}
\publisher{Zenodo}
\conference{The 19th Cambridge Workshop on Cool Stars, Stellar Systems, and the Sun}
\conferencedate{2016}

\title{How much can we trust high-resolution spectroscopic stellar atmospheric parameters?}
\author{
            S. Blanco-Cuaresma$^{1}$,
            T. Nordlander$^{2}$,
            U. Heiter$^{2}$,
            P. Jofr\'e$^{3}$,
            T. Masseron$^{3}$,
            L. Casamiquela$^{4}$,
            H. M. Tabernero$^{5}$,
            S. S. Bhat$^{6}$,
            A. R. Casey$^{3}$,
            J. Mel\'endez$^{7}$,
            and
            I. Ram\'irez$^{8}$
    }

\affiliation{
       $^{1}$ Observatoire de Gen\`eve, Universit\'e de Gen\`eve, CH-1290 Versoix, Switzerland \\
       $^{2}$ Department of Physics and Astronomy, Uppsala University, Box 516, 75120 Uppsala, Sweden \\
       $^{3}$ Institute of Astronomy, University of Cambridge, Madingley Road, Cambridge CB3 0HA, UK \\
       $^{4}$ Dep. F\'isica Qu\`antica i Astrof\'isica, Universitat de Barcelona, ICCUB-IEEC, Mart\'i Franqu\`es 1, E08028 Barcelona, Spain \\
       $^{5}$ Dpto. Astrof\'isica, Facultad de CC. F\'isicas, Universidad Complutense de Madrid, E-28040, Madrid, Spain \\
       $^{6}$ Christ University, Hosur road, Bangalore- 560029, India \\
       $^{7}$ Departamento de Astronomia do IAG/USP, Universidade de S\~ao Paulo; Rua do M\~atao 1226, S\~ao Paulo, 05508-900, SP, Brasil \\
       $^{8}$ McDonald Observatory and Department of Astronomy, University of Texas at Austin; 2515 Speedway, Stop C1402, Austin, TX 78712-1205, USA
}

\shorttitle{Trusting stellar atmospheric parameters}
\shortauthors{S. Blanco-Cuaresma et al. }

\abs{
The determination of atmospheric parameters depends on the use of radiative transfer codes (among other elements such as model atmospheres) to compute synthetic spectra and/or derive abundances from equivalent widths. However, it is common to mix results from different surveys/studies where different setups were used to derive the parameters. These inhomogeneities can lead us to inaccurate conclusions. In this work, we studied one aspect of the problem: When deriving atmospheric parameters from high-resolution stellar spectra, what differences originate from the use of different radiative transfer codes?

}

\begin{document}

\maketitle

\section{Introduction}

In the last years we have experienced a significant increase in the number of high-resolution stellar spectra available to the scientific community. This has been possible thanks to different surveys such as APOGEE \citep{2011AJ....142...72E} or the Gaia-ESO Public Spectroscopic Survey \citep[GES; ][]{2012Msngr.147...25G, 2013Msngr.154...47R}. The analysis of all these data can be carried out by different approaches like the synthetic spectral fitting technique or the classical equivalent width method. But such a huge quantity of data requires to automatize the analysis, and different authors have developed codes and pipelines to derive stellar atmospheric parameters \citep{2006MNRAS.370..141R, 2009A&A...501.1269K, 2012A&A...547A..13T, 2013A&A...558A..38M, 2013ApJ...766...78M, 2014MNRAS.443..698S, 2015ApJ...808...16N, 2015ApJ...812..128C, 2016ascl.soft05004M, 2016ApJS..223....8C, 2016arXiv160705792B}.

The result of all this recent development in the field of stellar spectroscopy is an increase in the number of available atmospheric parameters and chemical abundances, provided by independent studies and surveys using different setups (e.g., radiative transfer codes, stellar model atmospheres, continuum normalization). However, when combining all these results into a single data set in order to increase the statistical value, a challenge arises: The inhomogeneities of the original studies might end up affecting our scientific conclusions. Thus, it is important to evaluate the impact of these differences.  

Previous studies, such as \cite{2016arXiv160703130H}, already evaluated the global impact of different spectroscopic methods with different setups. In this study, we decided to tackle the problem by focusing on one single element. We fixed all the components of the spectroscopic analysis except one: the radiative transfer code. With this approach, we can better isolate the impact on the atmospheric parameter determination and compare the results between the synthetic spectral fitting technique and the equivalent width method. To address this complex experiment, we used iSpec\footnote{\url{http://www.blancocuaresma.com/s/}} \citep{2014A&A...569A.111B} and its flexibility to build a spectroscopic pipeline with different radiative transfer codes.

\section{Method}

iSpec is an open source spectroscopic framework that can be used to treat observed 1-D spectra (e.g., perform continuum normalization, measure and correct radial velocities) and it can also derive atmospheric parameters and abundances using the synthetic spectral fitting technique or the equivalent width method. However, until very recently, iSpec offered only one single radiative transfer code: SPECTRUM \citep{1994AJ....107..742G}.

In the stellar community there are plenty of other radiative transfer codes. Hence, we decided to integrate into iSpec some of the most popular ones:

\begin{itemize}
    \item WIDTH9/SYNTHE \citep{1993KurCD..18.....K, 2004MSAIS...5...93S}
    \item SME \citep{1996A&AS..118..595V}
    \item Turbospectrum \citep{1998A&A...330.1109A, 2012ascl.soft05004P}
    \item MOOG \citep{2012ascl.soft02009S}
\end{itemize}

To test all these codes, we used a set of very well known stars with reference parameters derived independently of spectroscopy: the Gaia FGK Benchmark Stars \citep{2014A&A...564A.133J, 2015A&A...582A..81J, 2015A&A...582A..49H, 2016A&A...592A..70H}. The spectra used in the experiment was obtained from its public high-resolution spectral library\footnote{\url{http://www.blancocuaresma.com/s/}} \citep{2014A&A...566A..98B}.

The synthetic spectral fitting technique implemented in iSpec does not use the full spectrum, but only the spectral features that the users choose (e.g., spectral regions that carry more information). The same happens for the equivalent width method due to the nature of the technique. To prepare the line selection, we used a solar spectrum from the library and we applied the following procedure for each radiative transfer code:

\begin{enumerate}
    \item Fix the atmospheric parameters to the solar reference values reported in \cite{2015A&A...582A..49H}.
    \item Normalize the spectrum, correct the radial velocity and convolve to a resolution of 47\,000 (using exactly the same algorithms and setup that will be used for the experiment).
    \item Automatically cross-match all the observed absorption lines with atomic data obtained from VALD \citep{2011BaltA..20..503K}.
    \item Derive chemical abundances for each of the identified lines using model atmospheres from MARCS\footnote{\url{http://marcs.astro.uu.se/}} \citep{2008A&A...486..951G} and solar abundances from \cite{2007SSRv..130..105G}.
    \item Select lines for which the determined chemical abundances is within $\pm$0.05 dex (given that it is a solar spectrum, we expect all [El/H] abundances to be close to 0.00 dex).
\end{enumerate}

Finally, we derive atmospheric parameters for all the Gaia FGK Benchmark Stars using only the absorption lines that were selected for all the radiative transfer codes in the solar spectrum (i.e., lines that all the codes are capable of correctly reproducing the solar spectrum).

\section{Results}

\begin{figure}
    \centering
    \includegraphics[width=0.85\linewidth]{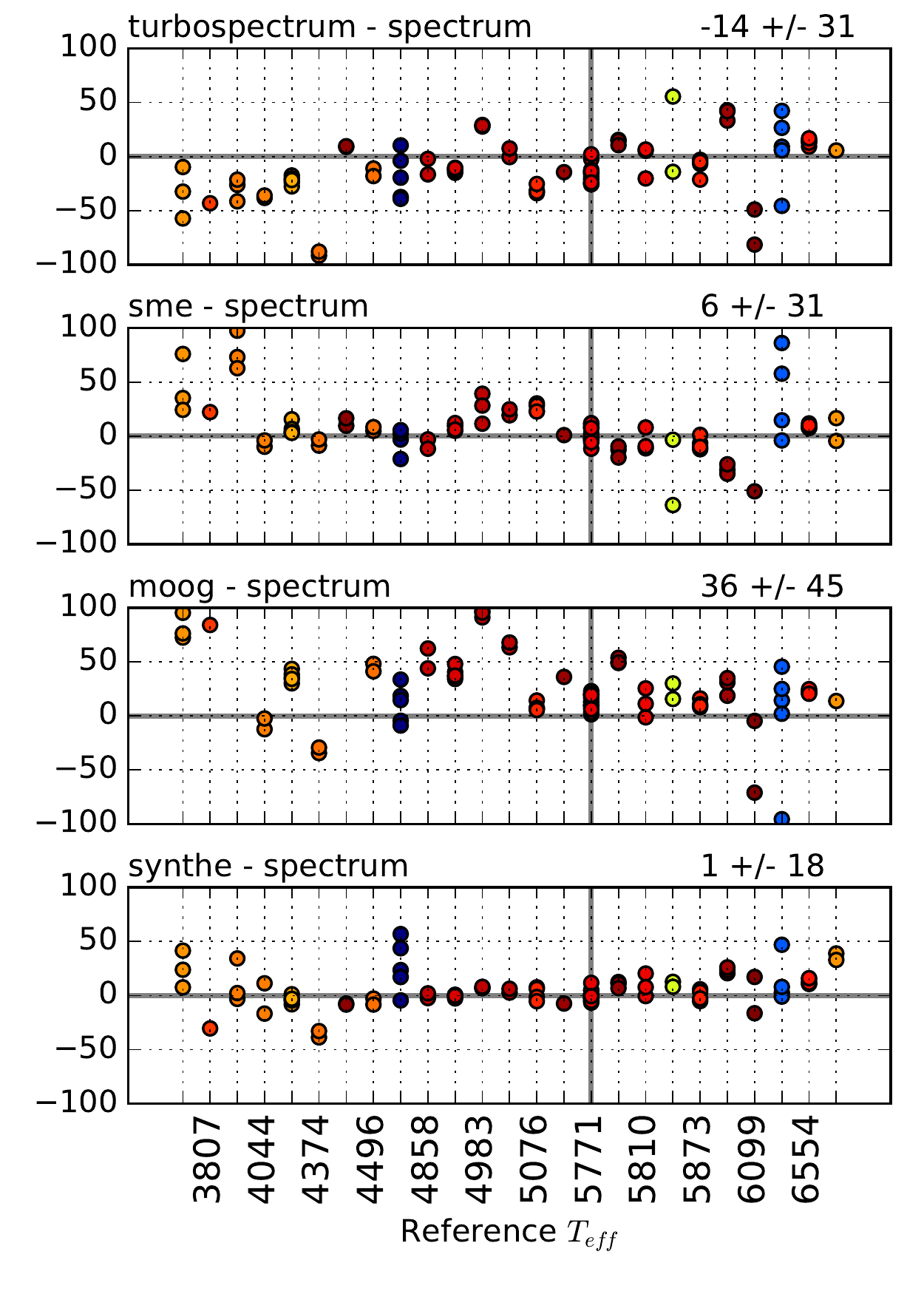}
    \caption{Differences in effective temperature (K) comparing all the radiative transfer codes (synthesis) against SPECTRUM (upper left text). Median difference and dispersion in the upper right text. Color code represents the star metallicity (blue: metal-poor; red: metal-rich).}
    \label{fig:synth_precision_teff}
\end{figure}

The results for the effective temperature when using the synthetic spectral fitting technique are shown in Fig.~\ref{fig:synth_precision_teff}. Each star has more than one result because several spectra were analyzed. The median difference and dispersion is shown in the upper right part of each subplot. The agreement between SPECTRUM and SYNTHE is remarkable, while Turbospectrum and SME is good although it deviates a little bit for colder and metal-poor stars. On the contrary, MOOG is the code that disagree the most with SPECTRUM or any of the other codes included in the study.

\begin{figure}
    \centering
    \includegraphics[width=0.85\linewidth]{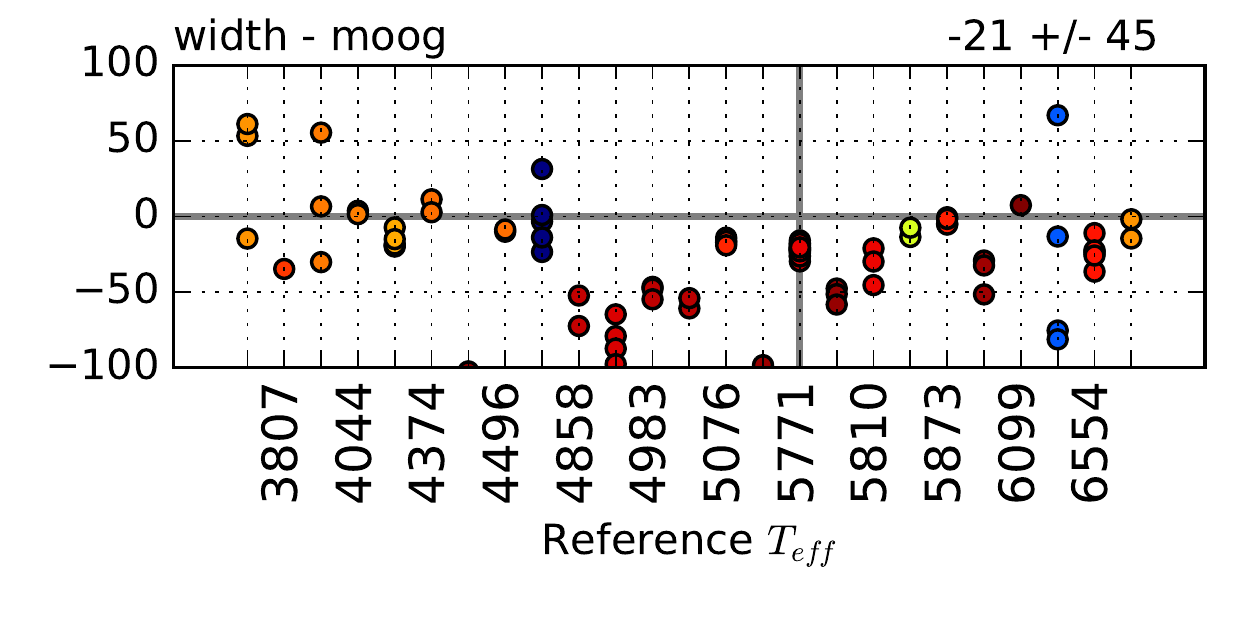}
    \caption{Differences in effective temperature (K) between two codes when using equivalent widths (upper left text). Median difference and dispersion in the upper right text. Color code as in Fig.~\ref{fig:synth_precision_teff}.}
    \label{fig:ew_precision_teff}
\end{figure}

If we check only the effective temperatures derived using the equivalent width method (Fig.~\ref{fig:ew_precision_teff}), we see that the level of disagreement (i.e. dispersion) between both methods is higher than the typical level for synthesis. 

\begin{figure}
    \centering
    \includegraphics[width=0.85\linewidth]{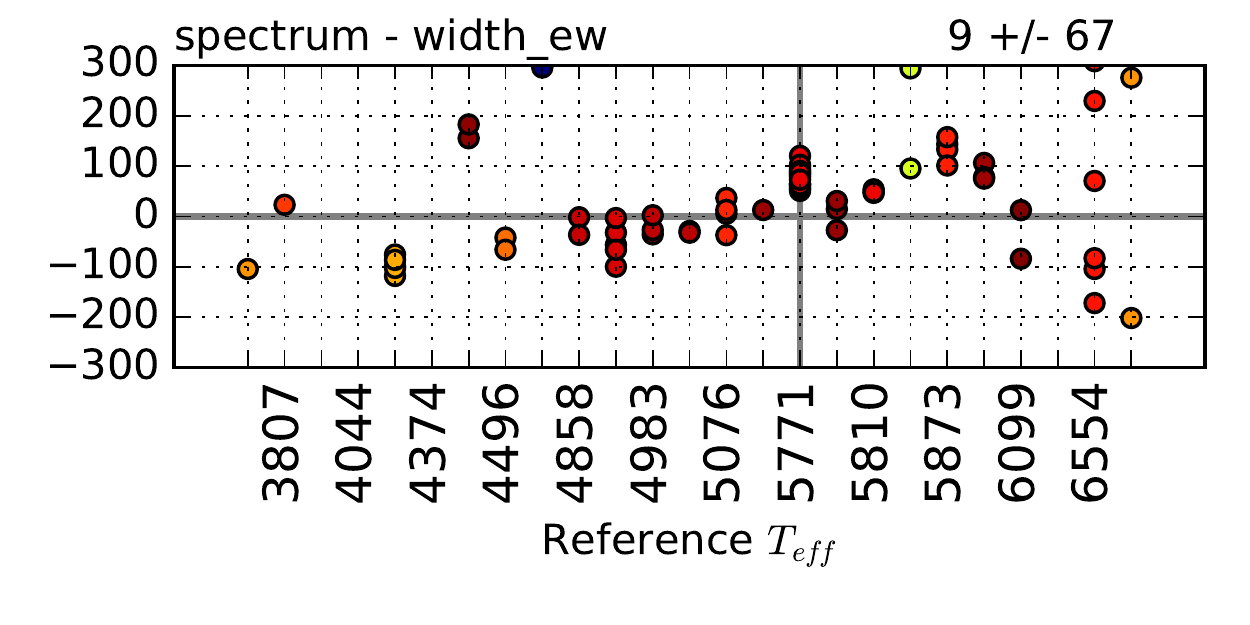}
    \caption{Differences in effective temperature (K) between codes using synthesis and equivalent width (upper left text). Median difference and dispersion in the upper right text. Color code as in Fig.~\ref{fig:synth_precision_teff}.}
    \label{fig:compare_precision_teff}
\end{figure}

Finally, when we compare the effective temperature derived using the spectral synthesis technique (with SPECTRUM) and the equivalent width method (with WIDTH9) as shown in Fig.~\ref{fig:compare_precision_teff}, the disagreement is even more important. Note that in the figure, the scale in the Y-axis was increased (compared to previous figures) to fit all the results.

For all these comparisons, the same global behaviour is found for surface gravities and metallicities. Codes that agree on effective temperature tend to agree on the rest of parameters and vice-versa.

\section{Conclusions}

We showed how different radiative transfer codes impact the determination of atmospheric parameters, and we quantified this impact for a wide range of stars. The level of agreement varies between different codes and it is noteworthy that the disagreement is higher when different methods are used (i.e. synthesis and equivalent width).

More importantly, this experiment was designed to keep all the variables fixed except the radiative transfer code and the method. The selection of lines was careful done to chose only the ones that reproduce better the Sun for all the codes. The condition were very favorable for a good converge toward similar values. Nevertheless, the disagreement cannot be completely ignored. This should discourage us from blindly mixing results coming from different sources where complete different setups and ingredients (and not only the radiative transfer codes) were used to derive their atmospheric parameters.

Additionally, the discrepancies in atmospheric parameters are going to propagate to the determination of chemical abundances. Thus, if we want accurate scientific conclusions when using abundances to study stellar aggregates and the Galaxy, it is necessary to make sure they were obtained homogeneously or, at least, perform an exhaustive assessment of the consequences of combining results obtain with different setups.

\section*{Acknowledgments}
{This work would not have been possible without the support of Laurent Eyer from the University of Geneva. UH and TN acknowledge support from the Swedish National Space Board (Rymdstyrelsen).}

\bibliographystyle{cs19proc}
\bibliography{SpectroscopicStellarAP}

\end{document}